\documentclass[12pt]{article}
\usepackage{amsmath}
\usepackage{amssymb}
\usepackage{fancyhdr}

\renewcommand{\maketitle}{
\begin{flushright} HCCSP-1002-06 \end{flushright}
\vskip2 true cm
    \begin{center}
      \Large
        {\bf Dirac Monopole from Lorentz Symmetry in
$N$-Dimensions:~~II. The Generalized Monopole}
        \vskip .3 true cm
      \normalsize
        Martin Land \\
        \vskip .3 true cm
        Department of Computer Science \\
        Hadassah College \\
        P. O. Box 1114, Jerusalem 91010, Israel
      \end{center}
      \vskip .5 true cm
}

\begin{document}

\title{}
\author{}
\maketitle

\begin{abstract}
In a previous paper, we found an extension of the $N$-dimensional Lorentz
generators that partially restores the closed operator algebra in the
presence of a Maxwell field, and is conserved under system evolution.
Generalizing the construction found by B\'{e}rard, Grandati, Lages and
Mohrbach for the angular momentum operators in the O(3)-invariant
nonrelativistic case, we showed that the construction can be maximally
satisfied in a three dimensional subspace of the full Minkowski space; this
subspace can be chosen to describe either the O(3)-invariant space sector,
or an O(2,1)-invariant restriction of spacetime. When the O(3)-invariant
subspace is selected, the field solution reduces to the Dirac monopole field
found in the nonrelativistic case. For the \hbox{O(2,1)-invariant} subspace,
the Maxwell field can be associated with a Coulomb-like potential of the
type $A^{\mu }(x)=n^{\mu }/\rho $, where $\rho =(x^{\mu }x_{\mu })^{1/2}$,
similar to that used by Horwitz and Arshansky to obtain a covariant
generalization of the hydrogen-like bound state. In this paper we elaborate
on the generalization of the Dirac monopole to $N$-dimensions.
\end{abstract}

\baselineskip7mm \parindent=0cm \parskip=10pt

\section{Introduction}

The Lorentz covariance of electrodynamics can be expressed through the
validity of the canonical commutation relations 
\begin{equation}
\lbrack x^{\mu },x^{\nu }]=0\mbox{\qquad\qquad}[p^{\mu },p^{\nu }]=0%
\mbox{\qquad\qquad}[x^{\mu },p^{\nu }]=-i\hbar g^{\mu \nu }  \label{eqn:0.10}
\end{equation}%
\begin{equation}
\left[ {x^{\mu },\,L^{\rho \lambda }}\right] =i\hbar \left( {x^{\lambda
}g^{\mu \rho }-x^{\rho }g^{\mu \lambda }}\right) \mbox{\qquad}\left[ {p^{\mu
},\,L^{\rho \lambda }}\right] =i\hbar \left( {g^{\mu \rho }p^{\lambda
}-g^{\mu \lambda }p^{\rho }}\right)  \label{eqn:0.20}
\end{equation}%
\begin{equation}
\left[ {L^{\mu \nu },\,L^{\lambda \rho }}\right] =i\hbar \left( {g^{\mu
\lambda }L^{\nu \rho }-g^{\mu \rho }L^{\nu \lambda }-g^{\nu \lambda }L^{\mu
\rho }+g^{\nu \rho }L^{\mu \lambda }}\right)  \label{eqn:0.30}
\end{equation}%
when the Lorentz generator is taken as 
\begin{equation}
L^{\mu \nu }=x^{\mu }p^{\nu }-x^{\nu }p^{\mu }  \label{eqn:0.40}
\end{equation}%
and the electromagnetic field is introduced through minimal coupling of a
gauge potential to the momentum%
\begin{equation}
p^{\mu }=m\dot{x}^{\mu }+eA^{\mu }.  \label{eqn:0.50}
\end{equation}%
Using (\ref{eqn:0.50}) to transform to the coordinate-velocity basis, the
relations (\ref{eqn:0.10}) become%
\begin{equation}
\lbrack x^{\mu },x^{\nu }]=0\mbox{\qquad\qquad}[x^{\mu },\dot{x}^{\nu
}]=-i\hbar g^{\mu \nu }  \label{eqn:0.60}
\end{equation}%
and 
\begin{equation}
\left[ \dot{x}^{\mu },\dot{x}^{\nu }\right] =\frac{1}{m^{2}}\left[ p^{\mu
}-eA^{\mu },p^{\nu }-eA^{\nu }\right] =-\dfrac{i\hbar e}{m^{2}}\left(
\partial ^{\mu }A^{\nu }-\partial ^{\nu }A^{\mu }\right) ,  \label{eqn:0.70}
\end{equation}%
a form that prompted Feynman \cite{dyson} to seek a ``
derivation''\ of Maxwell's equations assuming only the
commutation relations (\ref{eqn:0.60}) and noncommutivity of the velocities 
\begin{equation}
\left[ \dot{x}^{\mu },\dot{x}^{\nu }\right] =-\dfrac{i\hbar }{m^{2}}W^{\mu
\nu }\left( x\right) ,  \label{eqn:0.80}
\end{equation}%
without explicit recourse to canonical momentum, gauge potential, action, or
variational principle. It was later shown \cite{H-S} that assumptions (\ref%
{eqn:0.60}) are sufficiently strong to establish the self-adjointness of the
differential equations 
\begin{equation}
m\ddot{x}^{\mu }=F^{\mu }(\tau ,x,\dot{x}),  \label{eqn:0.90}
\end{equation}%
and it follows that this system is equivalent to a unique Lagrangian
mechanics \cite{Santilli} from which the full canonical system follows in
the standard manner. However, continuing in the spirit of Feynman's
`` naive''\ inquiry, we construct operators%
\begin{equation}
M^{\mu \nu }=m\left( x^{\mu }\dot{x}^{\nu }-x^{\nu }\dot{x}^{\mu }\right)
\label{eqn:0.100}
\end{equation}%
that, in light of (\ref{eqn:0.50}), are not generally equivalent to the
Lorentz generators (\ref{eqn:0.40}). The resulting commutation relations
among $x^{\mu }$, $\dot{x}^{\mu }$, and $M^{\mu \nu }$ contain terms that
depend on the field strength $W^{\mu \nu }\left( x\right) $, breaking the
Lie algebra for the Lorentz group.

In a previous paper \cite{part-I} we studied the generators (\ref{eqn:0.100}%
) in $N$ dimensions, and found an extension 
\begin{equation}
{\tilde{M}^{\mu \nu }=M^{\mu \nu }+Q^{\mu \nu }=m}\left( x^{\mu }\dot{x}%
^{\nu }-x^{\nu }\dot{x}^{\mu }\right) +x^{\mu }x_{\sigma }W^{\sigma \nu
}-x^{\nu }x_{\sigma }W^{\sigma \mu }-x^{\sigma }x_{\sigma }W{^{\mu \nu }}
\label{eqn:0.110}
\end{equation}%
that partially restores the closed algebra in the the coordinate-velocity
basis, without explicit reference to a gauge potential $A^{\mu }\left(
x\right) $. It was shown that the relations 
\begin{equation}
\left[ {x^{\mu },\,\tilde{M}^{\rho \lambda }}\right] =i\hbar \left( {%
x^{\lambda }g^{\mu \rho }-x^{\rho }g^{\mu \lambda }}\right) \mbox{\qquad}%
\left[ {\dot{x}^{\mu },\,\tilde{M}^{\rho \lambda }}\right] =i\hbar \left( {%
g^{\mu \rho }\dot{x}^{\lambda }-g^{\mu \lambda }\dot{x}^{\rho }}\right)
\label{eqn:0.120}
\end{equation}%
hold when the field is given by 
\begin{equation}
W^{\mu \nu }\left( x\right) =\frac{1}{{\left( {N-3}\right) !}}{\epsilon
^{\mu \nu \lambda _{0}\lambda _{1}\cdots \lambda _{N-3}}}\frac{x_{\lambda
_{0}}}{\left( x^{2}\right) ^{3/2}}U_{\lambda _{1}\cdots \lambda _{N-3}}
\label{eqn:0.130}
\end{equation}%
where ${U_{\lambda _{1}\cdots \lambda _{N-3}}}$ is totally antisymmetric,
and the dynamical evolution is restricted to the subspace 
\begin{equation}
x\left( \tau \right) \in x^{U}=\left\{ x~|~x^{\lambda _{1}}U_{\lambda
_{1}\lambda _{2}\cdots \lambda _{N-3}}=0\right\} .  \label{eqn:0.140}
\end{equation}%
The algebra of the generators was found to be 
\begin{equation}
\left[ {\tilde{M}^{\mu \nu },\,\tilde{M}^{\lambda \rho }}\right] =i\hbar
\left\{ {g^{\mu \lambda }\tilde{M}^{\nu \rho }-g^{\mu \rho }\tilde{M}^{\nu
\lambda }-g^{\nu \lambda }\tilde{M}^{\mu \rho }+g^{\nu \rho }\tilde{M}^{\mu
\lambda }}\right\} +\Delta _{2}^{\mu \nu \lambda \rho }  \label{eqn:0.150}
\end{equation}%
with 
\begin{equation}
\Delta _{2}^{\mu \nu \sigma \rho }=i\hbar \frac{1}{\left( x^{2}\right) ^{1/2}%
}\frac{1}{\left( N-3\right) !}\epsilon ^{\mu \sigma \rho \zeta \lambda
_{2}\cdots \lambda _{N-3}}x_{\zeta }g^{\nu \lambda _{1}}U_{\lambda
_{1}\lambda _{2}\cdots \lambda _{N-3}}  \label{eqn:0.160}
\end{equation}%
so that the symmetry-breaking term $\Delta _{2}^{\mu \nu \sigma \rho }$
vanishes for the three generators of O(3) or O(2,1) that leave the subspace $%
x^{U}$ invariant. The meaning of the electromagnetic field can be understood
in four dimensions, by taking the vector $U=\hat{t}$ along the time axis, so
that the field $W^{\mu \nu }$ describes a Dirac monopole, of the type
previously found in a nonrelativistic analysis by B\'{e}rard, Grandati,
Lages and Mohrbach \cite{BGLM}. Although the dynamical system that follows
from the commutation relations must be consistent with the gauge theory
posed in (\ref{eqn:0.10}) to (\ref{eqn:0.40}), the field $W^{\mu \nu }$,
found without reference to a gauge potential, describes a monopole solution
that does not follow from a standard gauge potential in a straightforward
manner. A related solution is found by taking the vector $U=\hat{z}$ along
the $z$-axis, for which the field $W^{\mu \nu }$describes an
O(2,1)-invariant solution associated with a potential of the type 
\begin{equation}
V\left( x\right) \sim \frac{1}{\sqrt{-t^{2}+\mathbf{x}^{2}}}
\end{equation}%
which may be seen as a relativistic generalization of the nonrelativistic
Coulomb force. A description of the relativistic bound state problem for the
scalar hydrogen atom was found \cite{bound} in the context of the
Horwitz-Piron \cite{H-P} formalism, using a potential of this form. In this
paper, we explore the higher dimensional Dirac monopole described in
expression (\ref{eqn:0.130}). The connection with the O(2,1)-invariant
generalization of the Coulomb force will be discussed in a subsequent paper.

\section{Gauge Theory from Commutation Relations}

\subsection{Stueckelberg-Lorentz force law}

According to Dyson's 1991 account \cite{dyson}, Feynman observed that posing
commutation relations of the form 
\begin{equation}
\left[ x^{i},x^{j}\right] =0\mbox{\qquad\qquad}m\,\left[ x^{i},\dot{x}^{j}%
\right] =i\hbar \,\delta ^{ij},  \label{eqn:0.170}
\end{equation}%
among the quantum operators for Euclidean position and velocity, where $\dot{%
x}^{i}=dx^{i}/dt$ and $i,j=1,2,3$, restricts the admissible forces in the
classical Newton's second law 
\begin{equation}
m\ddot{x}^{i}=F^{i}(t,x,\dot{x})  \label{eqn:0.180}
\end{equation}%
to the form 
\begin{equation}
m\ddot{x}^{i}=E^{i}(t,x)+\epsilon ^{ijk}\dot{x}_{j}H_{k}(t,x)
\label{eqn:0.190}
\end{equation}%
with fields that must satisfy 
\begin{equation}
\nabla \cdot \mathbf{H}=0\mbox{\qquad\qquad}\nabla \times \mathbf{E}+\frac{%
\partial }{\partial t}\mathbf{H}=0.  \label{eqn:0.200}
\end{equation}%
The velocity-dependent term in (\ref{eqn:0.190}) follows from the
`` naive''\ assumption that the velocities
have non-zero commutation relations, 
\begin{equation}
m^{2}\left[ \dot{x}^{i},\dot{x}^{j}\right] =-i\hbar F^{ij}(t,x)=-i\hbar
\epsilon ^{ijk}H_{k}(t,x),  \label{eqn:0.210}
\end{equation}%
so that Feynman's `` derivation''\ apparently
proceeds without explicit reference to canonical momentum, gauge potential,
action, or variational principle. It was later shown \cite{H-S} that the
assumptions (\ref{eqn:0.170}) are sufficiently strong to establish the
self-adjointness of the differential equations (\ref{eqn:0.180}), from which
it follows that this system is equivalent to a unique nonrelativistic
Lagrangian mechanics \cite{Santilli} with canonical momenta whose
relationship to the velocities leads directly to (\ref{eqn:0.210}). Several
authors observed \cite{conflict} that supposing Lorentz covariance in (\ref%
{eqn:0.180}) conflicts with the Euclidean assumptions in (\ref{eqn:0.170}),
and so (\ref{eqn:0.190}) cannot be interpreted as the Lorentz force in
Maxwell theory.

These results were generalized to the relativistic case \cite{Tanimura,
beyond} in curved $N$-dimensional spacetime by taking%
\begin{equation}
\lbrack x^{\mu },x^{\nu }]=0\mbox{\qquad\qquad}m[x^{\mu },\dot{x}^{\nu
}]=-i\hbar g^{\mu \nu }(x)\mbox{\qquad\qquad}\left[ \dot{x}^{\mu },\dot{x}%
^{\nu }\right] =\left( -\dfrac{i\hbar }{m^{2}}\right) f^{\mu \nu }(\tau ,x)
\label{eqn:0.220}
\end{equation}%
and 
\begin{equation}
m\ddot{x}^{\mu }=F^{\mu }(\tau ,x,\dot{x}).  \label{eqn:0.230}
\end{equation}%
where $\mu ,\nu =0,1,\cdots ,N-1$ and $x^{\mu }(\tau )$ and its derivatives
are functions of the Poincar\'{e}-invariant evolution parameter $\tau $.
Introducing the electric charge $e$ and a dimensional constant $\lambda $
required for consistency with Maxwell theory, the resulting system%
\begin{equation}
m[\ddot{x}^{\mu }+\Gamma ^{\mu \rho \nu }\dot{x}_{\rho }\dot{x}_{\nu
}]=\lambda e\left[ \epsilon ^{\mu }\left( \tau ,x\right) +f^{\mu \nu }\left(
\tau ,x\right) \dot{x}_{\nu }\right] ,  \label{eqn:0.240}
\end{equation}%
with usual affine connection 
\begin{equation}
\Gamma _{\mu \nu \rho }=\frac{1}{2}\left( \partial _{\rho }g_{\mu \nu
}+\partial _{\nu }g_{\mu \rho }-\partial _{\mu }g_{\nu \rho }\right)
\end{equation}%
$\ $and fields satisfying%
\begin{equation}
\partial _{\mu }f_{\nu \rho }+\partial _{\nu }f_{\rho \mu }+\partial _{\rho
}f_{\mu \nu }=0\mbox{\qquad}\partial _{\mu }\epsilon _{\nu }-\partial _{\nu
}\epsilon _{\mu }+{\frac{\partial }{\partial \tau }f_{\mu \nu }}=0,
\label{eqn:0.250}
\end{equation}%
is equivalent to the $(N+1)$-dimensional gauge theory associated with
Stueckelberg's relativistic mechanics \cite{Stueckelberg, saad}. The
commutation relations (\ref{eqn:0.220}) are defined with respect to a
covariant Heisenberg picture, in which operators evolve according to a
Poincar\'{e}-invariant parameter $\tau $. Thus, the position and velocity
operators $x^{\mu }(\tau )$ and $\dot{x}^{\mu }(\tau )$, for $\mu ,\nu
=0,1,\cdots ,N-1$, are the quantum analog of the classical quantities
introduced by Stueckelberg \cite{Stueckelberg} to construct his covariant
classical mechanics, in which the spacetime event $x^{\mu }\left( \tau
\right) $ traces out a general particle worldline as the parameter proceeds
monotonically from $\tau =-\infty $ to $\tau =\infty $.

\subsection{Pre-Maxwell field equations}

Formally extending the indices $\alpha ,\beta ,\gamma $ to %
\hbox{$\left(N+1\right)$-dimensions}, so that 
\begin{equation}
\mu ,\nu ,\rho =0,1,\cdots ,N-1\mbox{\qquad\qquad}\alpha ,\beta ,\gamma
=0,\cdots ,N
\end{equation}%
\begin{equation}
x^{N}=\tau \qquad \partial _{\tau }=\partial _{N}\qquad f_{\mu N}=-f_{N\mu
}=\epsilon _{\mu }  \label{eqn:0.260}
\end{equation}%
equations (\ref{eqn:0.240}) and (\ref{eqn:0.250}) become 
\begin{equation}
m[\ddot{x}^{\mu }+\Gamma ^{\mu \rho \nu }\dot{x}_{\rho }\dot{x}_{\nu
}]=\lambda e~f^{\mu \beta }(\tau ,x)\dot{x}_{\beta }  \label{eqn:0.270}
\end{equation}%
and 
\begin{equation}
\partial _{\alpha }f_{\beta \gamma }+\partial _{\beta }f_{\gamma \alpha
}+\partial _{\gamma }f_{\alpha \beta }=0.  \label{eqn:0.280}
\end{equation}%
The $N$ equations (\ref{eqn:0.270}) imply that%
\begin{equation}
\frac{d}{d\tau }(-\tfrac{1}{2}m\dot{x}^{\mu }\dot{x}_{\mu })=\lambda
e~f_{N\alpha }(\tau ,x)\dot{x}^{\alpha },  \label{eqn:0.290}
\end{equation}%
permitting the fields and particles to exchange mass. This classical
electrodynamics is equivalent \cite{beyond} to the Lagrangian system defined
by 
\begin{equation}
L=\tfrac{1}{2}M\dot{x}^{\mu }\dot{x}_{\mu }+\lambda e~a_{\mu }(\tau ,x)\dot{x%
}^{\mu }+\lambda e~a_{N}(\tau ,x)=\tfrac{1}{2}M\dot{x}^{\mu }\dot{x}_{\mu
}+\lambda e~a_{\alpha }(\tau ,x)\dot{x}^{\alpha }  \label{eqn:0.300}
\end{equation}%
with field strength related to to the gauge potential through%
\begin{equation}
f_{\alpha \beta }=\partial _{\alpha }a_{\beta }-\partial _{\beta }a_{\alpha
}.  \label{eqn:0.310}
\end{equation}%
Introducing in the action a kinetic term \cite{saad} for the $\tau $%
-dependent fields $f_{\alpha \beta }(\tau ,x)$ 
\begin{equation}
S=\int d\tau ~\left\{ \tfrac{1}{2}M\dot{x}^{\mu }\dot{x}_{\mu }+\lambda
ea_{\alpha }(\tau ,x)\dot{x}^{\alpha }\right\} -\tfrac{\lambda }{4}\int
d^{N-1}x~d\tau ~f^{\alpha \beta }(\tau ,x)f_{\alpha \beta }(\tau ,x)
\label{eqn:0.320}
\end{equation}%
leads to the inhomogeneous source equation 
\begin{equation}
\partial _{\beta }f^{\alpha \beta }(\tau ,x)=j^{\alpha }(\tau ,x)
\label{eqn:0.330}
\end{equation}%
with classical current of the form%
\begin{equation}
j^{\alpha }(\tau ,x)=e\dot{z}^{\alpha }(\tau )\ \delta ^{N-1}\left(
x-z\right) .  \label{eqn:0.340}
\end{equation}%
The associated quantum theory 
\begin{equation}
\Bigl[i\partial _{\tau }+\lambda ea_{N}\Bigr]\psi (x,\tau )=\frac{1}{2M}%
\Bigl[p^{\mu }-\lambda ea^{\mu }\Bigr]\Bigl[p_{\mu }-\lambda ea_{\mu }\Bigr]%
\psi (x,\tau ),  \label{eqn:0.350}
\end{equation}%
proposed by Sa'ad, Horwitz and Arshansky \cite{saad}, is invariant under
local gauge transformations of the type 
\begin{equation}
\psi (x,\tau )\rightarrow e^{i\lambda e\Lambda (x,\tau )}\psi (x,\tau )%
\mbox{\qquad}\mbox{\qquad}a_{\alpha }(x,\tau )\rightarrow a_{\alpha }(x,\tau
)+\partial _{\alpha }\Lambda (x,\tau ).
\end{equation}%
Global gauge symmetry leads to an $\left( N+1\right) $-dimensional conserved
current 
\begin{equation}
\partial _{\alpha }j^{\alpha }(\tau ,x)=\partial _{\mu }j^{\mu }+\partial
_{\tau }j^{N}=0
\end{equation}%
\smallskip where%
\begin{equation}
j^{\mu }(x,\tau )=\frac{-ie}{2M}\Bigl[\psi ^{\ast }(\partial ^{\mu
}-i\lambda ea^{\mu })\psi -\psi (\partial ^{\mu }+i\lambda ea^{\mu })\psi
^{\ast }\Bigr]\mbox{\qquad}j^{N}(x,\tau )=e\Bigl|\psi (x,\tau )\Bigr|^{2}
\label{eqn:0.360}
\end{equation}%
so that $\Bigl|\psi (x,\tau )\Bigr|^{2}$ is interpreted as the
positive-definite probability density at $\tau $ of finding the event at the
spacetime point $x$. Equations (\ref{eqn:0.280}) and (\ref{eqn:0.330}) are
formally identical to the standard Maxwell equations, but differ in two
significant ways --- first, these are $\left( N+1\right) $-dimensional
equations in $N$-dimensions, and second, the structure of the source current
(\ref{eqn:0.340}) depends directly on the instantaneous event $x^{\mu }(\tau
)$, while the Maxwell current has support on the entire worldline%
\begin{equation}
J^{\mu }(x)=e\int d\tau ~\dot{z}^{\mu }(\tau )\ \delta ^{N-1}\left(
x-z\right) .
\end{equation}%
Standard Maxwell theory can be recovered from the Stueckelberg theory as an
equilibrium limit, defined pointwise in $x$ as $\tau \rightarrow \pm \infty $
by the conditions 
\begin{equation}
f_{N\mu }(\tau ,x)=0\qquad \mathrm{and}\qquad \partial _{\tau }f^{\mu \nu
}(\tau ,x)=0.
\end{equation}%
Integrating (\ref{eqn:0.280}) and (\ref{eqn:0.330}) over $\tau $
concatenates events into worldlines \cite{concat}, recovering 
\begin{equation}
\partial _{\mu }F_{\nu \rho }+\partial _{\nu }F_{\rho \mu }+\partial _{\rho
}F_{\mu \nu }=0\qquad \qquad \partial _{\nu }F^{\mu \nu }=J^{\mu }
\end{equation}%
where 
\begin{equation}
F^{\mu \nu }(x)=\int_{-\infty }^{\infty }d\tau \;f^{\mu \nu }(x,\tau )%
\mbox{\quad}\mathrm{and}\mbox{\quad}J^{\mu }(x)=\int_{-\infty }^{\infty
}d\tau \;j^{\mu }(x,\tau ).
\end{equation}%
It follows that $\lambda $ has dimensions of time. Thus the instantaneous
pre-Maxwell field is written as $f^{\mu \nu }(x,\tau )$, and Maxwell-like
fields, obtained by concatenation or by direct solution of Maxwell equations
in $N$-dimensions, are written as $F^{\mu \nu }(x)$. In the remainder of
this paper, we will consider only $\tau $-independent Maxwell-like fields.

\section{Electromagnetic Duality and the Monopole}

The standard treatment of the Dirac monopole in four dimensional spacetime
is deceptively simple. By introducing a magnetic current vector, the
magnetic field acquires a non-zero divergence (source), and the Maxwell
equations become symmetric under exchange of the electric and magnetic
sectors, so that any electric field solution induced by an electric current
can be associated with a magnetic field solution induced by a magnetic
current. However, this exchange symmetry is an artifact of four dimensional
spacetime. Ignoring the laboratory origins of electromagnetics and viewing
the Maxwell theory abstractly, as a pair of Lorentz covariant differential
equations for a second rank tensor field on $N$-dimensional spacetime, there
are two alternative approaches to the introduction of a second source
current --- in four dimensions, these approaches become equivalent.

The first approach, in analogy to the historical route, follows from the
non-covariant decomposition of the tensor field, in some particular time +
space reference frame, into an electric field associated with a source and a
sourceless magnetic field. In this approach, a source may be introduced for
the magnetic field, but this current does not make the Maxwell theory
symmetric under exchange of the electric and magnetic sectors, except in the
well-known case of $N=4$. Moreover, for $N\neq 4,$ the exchange of these
non-covariant sectors is not equivalent to a duality transformation, and the
tensor Maxwell equations are not duality-symmetric.

The second approach generalizes the Maxwell tensor field to a Clifford
number field, leading to duality symmetric field equations. This approach
provides an alternative description of the magnetic monopole, based on a
covariant decomposition of the field into symmetric sectors, however, only
in $N=4$ can the field sectors exchanged by this duality transformation be
identified with the field sectors found by non-covariant decomposition. In
this section, we will show that the generalized magnetic monopole found in 
\cite{part-I} supports approach of the duality invariance.

\subsection{Clifford algebra formulation}

The spacetime algebra formalism \cite{clifford} provides a high degree of
notational compactness by representing the usual tensorial objects of
physics as index-free elements in a Clifford algebra. The formalism
considerably simplifies the treatment of the magnetic monopole in higher
dimensions because it eases the transition between covariant and space +
time points of view, and contains a natural geometric representation of the
duality transformation. In Clifford algebra, the product of two vectors
separates naturally into a symmetric part and antisymmetric part 
\begin{equation}
ab=\tfrac{1}{2}\left( {ab+ba}\right) +\tfrac{1}{2}\left( {ab-ba}\right)
=a\cdot b+a\wedge b
\end{equation}%
where the symmetric part is identified with the scalar inner product, and
the rank 2 antisymmetric part is called a bivector. The general Clifford
number is a direct sum of multivectors of rank $0,1,\ldots ,N$ 
\begin{eqnarray}
A &=&A_{0}+A_{1}+A_{2}+A_{3}+\cdots +A_{N} \\
&=&A_{0}+A_{1}^{\alpha }{\mathbf{e}}_{\alpha }+\tfrac{1}{2}A_{2}^{\alpha
\beta }{\mathbf{e}}_{\alpha }\wedge {\mathbf{e}}_{\beta }+\cdots +\tfrac{1}{%
N!}A_{N}^{\alpha _{0}\alpha _{2}\cdots \alpha _{N-1}}{\mathbf{e}}_{\alpha
_{0}}\wedge {\mathbf{e}}_{\alpha _{1}}\wedge \cdots \wedge {\mathbf{e}}%
_{\alpha _{N-1}}
\end{eqnarray}%
expanded on the $2^{N}$-dimensional basis 
\begin{equation}
\left\{ {1,{\mathbf{e}}_{\alpha },{\mathbf{e}}_{\alpha }\wedge {\mathbf{e}}%
_{\beta },{\mathbf{e}}_{\alpha }\wedge {\mathbf{e}}_{\beta }\wedge {\mathbf{e%
}}_{\gamma },\cdots ,{\mathbf{e}}_{0}\wedge {\mathbf{e}}_{1}\wedge \cdots
\wedge {\mathbf{e}}_{N-1}}\right\} .
\end{equation}%
The most important algebraic rules are%
\begin{eqnarray}
aA_{r} &=&a\left( {a_{1}\wedge a_{2}\wedge \cdots \wedge a_{r}}\right)
=a\cdot A_{r}+a\wedge A_{r} \\
a\cdot A_{r} &=&\sum\nolimits_{k=1}^{r}{\left( {-1}\right) ^{k+1}\left( {%
a\cdot a_{k}}\right) ~}a_{1}\wedge \cdots \wedge a_{k-1}\wedge a_{k+1}\wedge
\cdots \wedge a_{r} \\
a\wedge A_{r} &=&a\wedge a_{1}\wedge a_{2}\wedge \cdots \wedge a_{r} \\
\left( a\wedge b\right) \cdot A_{r} &=&a\cdot \left( b\cdot A_{r}\right) \\
\mathbf{i} &=&{\mathbf{e}}_{0}\wedge {\mathbf{e}}_{1}\wedge \cdots \wedge {%
\mathbf{e}}_{N-1}={\mathbf{e}}_{0}{\mathbf{e}}_{1}\cdots {\mathbf{e}}_{N-1}
\label{eqn:0.370} \\
\mathbf{i}^{2} &=&\left( {-1}\right) ^{\frac{{N\left( {N-1}\right) }}{2}%
}\det \left( g_{ij}\right)  \label{eqn:0.380} \\
\mathbf{i}\left[ {{\mathbf{e}}_{\alpha _{1}}\wedge \cdots \wedge {\mathbf{e}}%
_{\alpha _{r}}}\right] &=&g_{\alpha _{1}\alpha _{1}}\cdots g_{\alpha
_{r}\alpha _{r}}\tfrac{1}{\left( N-r\right) {!}}{\ \epsilon ^{\alpha
_{1}\cdots \alpha _{r}\alpha _{r+1}\cdots \alpha _{N}}\left[ {\ {\mathbf{e}}%
_{\alpha _{r+1}}\wedge \cdots \wedge {\mathbf{e}}_{\alpha _{N}}}\right] }
\label{eqn:0.390} \\
a\cdot \left( \mathbf{i}{A_{r}}\right) &=&\left( {-1}\right) ^{N-1}\mathbf{i}%
\left( {a\wedge A_{r}}\right)  \label{eqn:0.400} \\
a\wedge \left( \mathbf{i}{A_{r}}\right) &=&\left( {-1}\right) ^{N-1}\mathbf{i%
}\left( {a\cdot A_{r}}\right) .  \label{eqn:0.410}
\end{eqnarray}%
We choose the flat metric%
\begin{equation}
g_{\alpha \beta }={\mathbf{e}}_{\alpha }\cdot {\mathbf{e}}_{\beta }=\mathrm{%
diag}\left( -1,1,\cdots ,1\right)
\end{equation}%
for which the unit pseudoscalar $\mathbf{i}$ satisfies%
\begin{equation}
\mathbf{i}^{2}=\left( {-1}\right) ^{\frac{{N\left( {N-1}\right) }}{2}}\det
\left( g_{ij}\right) =-\left( {-1}\right) ^{\frac{{N\left( {N-1}\right) }}{2}%
}.
\end{equation}%
In the Clifford algebra formulation, Maxwell's equations in $N$-dimensions
are%
\begin{equation}
dF=-J  \label{eqn:0.420}
\end{equation}%
in which the electromagnetic field strength bivector (antisymmetric second
rank tensor) is%
\begin{equation}
F=\dfrac{1}{2}F^{\alpha \beta }\left( \mathbf{e}_{\alpha }\wedge \mathbf{e}%
_{\beta }\right) ,
\end{equation}%
and the gradient and current vectors are $d=\partial ^{\alpha }\mathbf{e}%
_{\alpha }$ and $J=J^{\alpha }\mathbf{e}_{\alpha }$. The LHS of (\ref%
{eqn:0.420}) separates into the divergence and exterior derivative 
\begin{equation}
d\cdot F+d\wedge F=-J
\end{equation}%
so that associating terms of equal rank rank leads to%
\begin{eqnarray}
&&d\cdot F=-J  \label{eqn:0.430} \\
&&d\wedge F=0.  \label{eqn:0.440}
\end{eqnarray}%
Equation (\ref{eqn:0.430}) expresses the inhomogeneous Maxwell equations, as
seen from the component tensor form%
\begin{equation}
d\cdot F=\partial ^{\alpha }\left( \dfrac{1}{2}F^{\beta \gamma }\right) 
\mathbf{e}_{\alpha }\cdot \left( \mathbf{e}_{\beta }\wedge \mathbf{e}%
_{\gamma }\right) =\partial ^{\alpha }F^{\beta \gamma }\dfrac{1}{2}\left(
g_{\alpha \beta }\mathbf{e}_{\gamma }-g_{\alpha \gamma }\mathbf{e}_{\beta
}\right) =\left( \partial _{\beta }F^{\beta \gamma }\right) \mathbf{e}%
_{\gamma }.
\end{equation}%
Similarly, using the total antisymmetry of $\mathbf{e}_{\alpha }\wedge 
\mathbf{e}_{\beta }\wedge \mathbf{e}_{\gamma }$ we expand (\ref{eqn:0.440})
as 
\begin{equation}
d\wedge F=\partial ^{\alpha }\left( \dfrac{1}{2}F^{\beta \gamma }\right) 
\mathbf{e}_{\alpha }\wedge \mathbf{e}_{\beta }\wedge \mathbf{e}_{\gamma }=%
\dfrac{1}{3!}\left( \partial ^{\alpha }F^{\beta \gamma }+\partial ^{\beta
}F^{\gamma \alpha }+\partial ^{\gamma }F^{\alpha \beta }\right) \mathbf{e}%
_{\alpha }\wedge \mathbf{e}_{\beta }\wedge \mathbf{e}_{\gamma }=0,
\label{eqn:0.450}
\end{equation}%
expressing the homogeneous Maxwell equations. The wave equation and current
conservation follow from (\ref{eqn:0.420}) as%
\begin{equation}
d^{2}F=-dJ=-d\cdot J-d\wedge J,
\end{equation}%
which separates by rank into%
\begin{equation}
d^{2}F=-d\wedge J\mbox{\qquad}d\cdot J=0.
\end{equation}%
We will refer to the field equations in the form (\ref{eqn:0.430}) and (\ref%
{eqn:0.440}) as the Jacobi point of view, because the exterior derivative in
(\ref{eqn:0.450}) follows from the Jacobi identity for the commutator $\left[
\dot{x}^{\alpha },\left[ \dot{x}^{\beta },\dot{x}^{\gamma }\right] \right] $%
. On the other hand, writing the dual to $F$ as%
\begin{equation}
\tilde{F}={\mathbf{i}}F  \label{eqn:0.455}
\end{equation}%
relation (\ref{eqn:0.400}) permits equation (\ref{eqn:0.440}) to be written 
\begin{equation}
\left( -1\right) ^{N-1}{\mathbf{i}}\left( d\wedge F\right) =d\cdot \left( {%
\mathbf{i}}F\right) =d\cdot \tilde{F}=0,  \label{eqn:0.460}
\end{equation}%
and in what we will call the divergence point of view, Maxwell theory can be
represented as a pair of inequivalent tensor structures, whose divergences
are associated with a source (or sourcelessness) through%
\begin{equation*}
d\cdot F=J\mbox{\qquad}\mbox{\qquad}d\cdot \tilde{F}=0.
\end{equation*}%
In this point of view, the divergenceless of $\tilde{F}$ expresses the
asymmetry of the Maxwell theory under duality --- the exchange of $F$ and $%
\tilde{F}$ --- and is related to the absence of a source for the magnetic
field in four dimensions. Thus, a theory describing magnetic monopoles with
no electric charges would appear in the divergence point of view as%
\begin{equation}
d\cdot G=0\mbox{\qquad}\mbox{\qquad}d\cdot \tilde{G}=\tilde{J}^{\left(
m\right) }
\end{equation}%
and in the Jacobi point of view%
\begin{equation}
d\cdot G=0\mbox{\qquad}\mbox{\qquad}d\wedge G=J^{\left( m\right) }
\label{eqn:0.465}
\end{equation}%
with trivector current $J^{\left( m\right) }$.

\subsection{Non-manifestly covariant field equations}

The usual distinction between the electric and magnetic fields is based on
the decomposition of the manifestly covariant field equations into a time +
space formulation in some reference frame. Choosing a time direction $%
\mathbf{e}_{0}$, the field strength separates into time and space components
as%
\begin{equation}
F=\dfrac{1}{2}F^{\alpha \beta }\left( \mathbf{e}_{\alpha }\wedge \mathbf{e}%
_{\beta }\right) =F^{0i}\left( \mathbf{e}_{0}\wedge \mathbf{e}_{i}\right) +%
\dfrac{1}{2}F^{ij}\left( \mathbf{e}_{i}\wedge \mathbf{e}_{j}\right) =\mathbf{%
e}_{0}\wedge \mathbf{E}+\mathbf{F}  \label{eqn:0.470}
\end{equation}%
where the vector $\mathbf{E}$ and the bivector $\mathbf{F}$, defined as%
\begin{equation}
\mathbf{E}=F^{0i}\mathbf{e}_{i}\mbox{\qquad}\mathbf{F}=\dfrac{1}{2}%
F^{ij}\left( \mathbf{e}_{i}\wedge \mathbf{e}_{j}\right) \mbox{\qquad}%
i,j=1,...,N-1,
\end{equation}%
have only space components. In this reference frame, the classical Lorentz
force decomposes as 
\begin{equation}
m\frac{{d^{2}}x}{{d\tau ^{2}}}=eF\cdot \dot{x}=e\left( \mathbf{e}_{0}\wedge 
\mathbf{E}\right) \cdot \dot{x}+e\mathbf{F}\cdot \dot{x}=e\left( \mathbf{E}%
\cdot \dot{x}\right) \mathbf{e}_{0}-e\dot{x}_{0}\mathbf{E}+e\mathbf{F}\cdot 
\mathbf{\dot{x}}
\end{equation}%
with time and space components%
\begin{equation}
m\frac{{d^{2}}x_{0}}{{d\tau ^{2}}}=e\left( \mathbf{E}\cdot \dot{\mathbf{x}}%
\right) \mbox{\qquad\qquad\qquad}m\frac{{d^{2}}\mathbf{x}}{{d\tau ^{2}}}=e%
\dot{x}^{0}\mathbf{E}+e\mathbf{F}\cdot \dot{\mathbf{x}}.  \label{eqn:0.490}
\end{equation}%
Expressions (\ref{eqn:0.490}) distinguish the roles of $\mathbf{E}$ and $%
\mathbf{F}$, showing that the electric-like force $e\dot{x}^{0}\mathbf{E}$
can perform work on a test particle and may be nonzero in a co-moving frame,
while the magnetic-like force $e\mathbf{F}\cdot \dot{\mathbf{x}}$ vanishes
in the co-moving frame, and is seen from%
\begin{equation}
\left[ \mathbf{F}\cdot \dot{\mathbf{x}}\right] \cdot \dot{\mathbf{x}}=%
\mathbf{F}\cdot \left[ \dot{\mathbf{x}}\wedge \dot{\mathbf{x}}\right] =0
\end{equation}%
to be orthogonal to the velocity. The analog of the three-vector Maxwell
equations are found by decomposing the $N$-divergence into%
\begin{equation}
d=\partial ^{0}\mathbf{e}_{0}+\nabla \mbox{\qquad\qquad\qquad}\nabla
=\partial ^{i}\mathbf{e}_{i}  \label{eqn:0.500}
\end{equation}%
and applying (\ref{eqn:0.500}) to (\ref{eqn:0.470}) for the inhomogeneous
equation%
\begin{equation}
d\cdot F=d\cdot \left( \mathbf{e}_{0}\wedge \mathbf{E}+\mathbf{F}\right)
=\partial _{0}\mathbf{E}-\nabla \cdot \mathbf{Ee}_{0}+\nabla \cdot \mathbf{F}%
=-\left( \rho \mathbf{e}_{0}+\mathbf{J}\right)
\end{equation}%
which separates into the time and space components%
\begin{equation}
\nabla \cdot \mathbf{E}=\rho \mbox{\qquad\qquad\qquad}-\nabla \cdot \mathbf{F%
}-\partial _{0}\mathbf{E}=\mathbf{J}.  \label{eqn:0.520}
\end{equation}%
We expand the unit pseudoscalar in the time + space reference frame as%
\begin{equation}
{\mathbf{i}}=\mathbf{e}_{0}{\mathbf{i}}_{\left( space\right) }\mbox{\qquad}{%
\mathbf{i}}_{\left( space\right) }=\mathbf{e}_{1}...\mathbf{e}_{N-1}%
\mbox{\qquad}{\mathbf{i}}_{\left( space\right) }^{2}=\left( -1\right) ^{%
\frac{\left( N-1\right) \left( N-2\right) }{2}}
\end{equation}%
and notice that the taking the dual of the field strength%
\begin{eqnarray}
\tilde{F} &=&{\mathbf{i}}F={\mathbf{i}}\mathbf{e}_{0}\wedge \mathbf{E}+{%
\mathbf{i}}\mathbf{F}  \notag \\
&=&\mathbf{e}_{0}{\mathbf{i}}_{\left( space\right) }\mathbf{e}_{0}\mathbf{E}+%
\mathbf{e}_{0}{\mathbf{i}}_{\left( space\right) }\mathbf{F}  \notag \\
&=&\mathbf{e}_{0}\wedge \left[ {\mathbf{i}}_{\left( space\right) }\mathbf{F}%
\right] +\left( -1\right) ^{N}\left[ {\mathbf{i}}_{\left( space\right) }%
\mathbf{E}\right]  \label{eqn:0.540}
\end{eqnarray}%
exchanges the roles of the electric-like vector and the magnetic-like
bivector. Writing 
\begin{equation}
\mathbf{\tilde{F}}=-{\mathbf{i}}_{\left( space\right) }\mathbf{F}
\label{eqn:0.545}
\end{equation}%
and using (\ref{eqn:0.400}) and (\ref{eqn:0.410}), the time + space theory,
in analogy to the three-vector Maxwell equations, is expressed as%
\begin{equation}
\begin{array}{lrr}
\nabla \cdot \mathbf{E}=\rho & \mbox{\hspace{30 pt}} & \left[ \left(
-1\right) ^{\frac{\left( N-1\right) \left( N-2\right) }{2}+N}\right] {%
\mathbf{i}}_{\left( space\right) }\nabla \wedge \mathbf{\tilde{F}}-\partial
_{0}\mathbf{E}=\mathbf{J~~}%
\end{array}
\label{eqn:0.546a}
\end{equation}%
\begin{equation}
\begin{array}{lrr}
\nabla \cdot \mathbf{\tilde{F}}=0 & \mbox{\hspace{115 pt}} & -{\mathbf{i}}%
_{\left( space\right) }\nabla \wedge \mathbf{E}+\partial _{0}\mathbf{\tilde{F%
}}=0.%
\end{array}
\label{eqn:0.546b}
\end{equation}%
In this form, the divergencelessness of the field $\tilde{F}$ is equivalent
to the sourcelessness of the field $\mathbf{\tilde{F}}$.

\subsection{Duality and the Dirac monopole in $N=4$ dimensions}

The standard treatment of the magnetic monopole relies on two closely
related simplifications of Maxwell theory that obtain only in four
dimensions. First, the $\left( N-1\right) $-component electric-like vector $%
\mathbf{E}$ and the $\frac{\left( N-1\right) \left( N-2\right) }{2}$%
-component magnetic-like bivector $\mathbf{F}$ can only have an equal number
of degrees of freedom in the special case that%
\begin{equation}
\left( N-1\right) =\frac{\left( N-1\right) \left( N-2\right) }{2}\qquad
\Rightarrow \qquad N=4.
\end{equation}%
Then, the Maxwell field strength tensor%
\begin{eqnarray}
F &=&\left[ F^{01}\left( \mathbf{e}_{0}\wedge \mathbf{e}_{1}\right)
+F^{02}\left( \mathbf{e}_{0}\wedge \mathbf{e}_{2}\right) +F^{03}\left( 
\mathbf{e}_{0}\wedge \mathbf{e}_{3}\right) \right.  \notag \\
&&\mbox{\qquad\qquad}\left. +F^{12}\left( \mathbf{e}_{1}\wedge \mathbf{e}%
_{2}\right) +F^{13}\left( \mathbf{e}_{1}\wedge \mathbf{e}_{3}\right)
+F^{23}\left( \mathbf{e}_{2}\wedge \mathbf{e}_{3}\right) \right]
\label{eqn:0.550}
\end{eqnarray}%
decomposes into the non-manifestly covariant form%
\begin{eqnarray}
F &=&\mathbf{e}_{0}\wedge \mathbf{E}+\mathbf{F}  \notag \\
&=&\left[ E^{1}\left( \mathbf{e}_{0}\wedge \mathbf{e}_{1}\right)
+E^{2}\left( \mathbf{e}_{0}\wedge \mathbf{e}_{2}\right) +E^{3}\left( \mathbf{%
e}_{0}\wedge \mathbf{e}_{3}\right) \right. \\
&&\mbox{\qquad\qquad}\left. +H^{3}\left( \mathbf{e}_{1}\wedge \mathbf{e}%
_{2}\right) -H^{2}\left( \mathbf{e}_{1}\wedge \mathbf{e}_{3}\right)
+H^{1}\left( \mathbf{e}_{2}\wedge \mathbf{e}_{3}\right) \right] ,
\end{eqnarray}%
allowing the magnetic-like bivector $F$ to be identified with a magnetic
vector $\mathbf{H}$ through%
\begin{eqnarray}
\mathbf{F} &=&H^{1}\left( \mathbf{e}_{2}\wedge \mathbf{e}_{3}\right)
-H^{2}\left( \mathbf{e}_{1}\wedge \mathbf{e}_{3}\right) +H^{3}\left( \mathbf{%
e}_{1}\wedge \mathbf{e}_{2}\right) \\
&=&\mathbf{e}_{1}\mathbf{e}_{2}\mathbf{e}_{3}\left[ H^{1}\mathbf{e}_{1}+H^{2}%
\mathbf{e}_{2}+H^{3}\mathbf{e}_{3}\right] \\
&=&{\mathbf{i}}_{\left( space\right) }\mathbf{H}.  \label{eqn:0.555}
\end{eqnarray}%
Applying definition (\ref{eqn:0.545}) to (\ref{eqn:0.555}) we recover the
identification 
\begin{equation}
\mathbf{\tilde{F}}=-{\mathbf{i}}_{\left( space\right) }\left[ {\mathbf{i}}%
_{\left( space\right) }\mathbf{H}\right] =\mathbf{H}
\end{equation}%
and by recognizing that in three-space%
\begin{equation}
a\wedge b=a^{i}b^{j}\left( \mathbf{e}_{i}\wedge \mathbf{e}_{j}\right)
=a_{i}b_{j}\left[ \epsilon ^{ijk}{\mathbf{i}}_{\left( space\right) }\mathbf{e%
}_{k}\right] ={\mathbf{i}}_{\left( space\right) }\left( \epsilon
^{ijk}a_{i}b_{j}\right) \mathbf{e}_{k}={\mathbf{i}}_{\left( space\right)
}a\times b,  \label{eqn:0.557}
\end{equation}%
the Maxwell equations (\ref{eqn:0.546a}) and (\ref{eqn:0.546b}) assume the
standard 3-vector form%
\begin{eqnarray}
\nabla \cdot \mathbf{E} &=&\rho \mbox{\qquad}\nabla \times \mathbf{H}-\dfrac{%
\partial }{\partial t}\mathbf{E}=\mathbf{J}  \label{eqn:0.554} \\
\nabla \cdot \mathbf{H} &=&0\mbox{\qquad}\nabla \times \mathbf{E}+\dfrac{%
\partial }{\partial t}\mathbf{H}=0.  \label{eqn:0.556}
\end{eqnarray}%
The second simplification follows from (\ref{eqn:0.390}) which shows that
the rank of $\tilde{F}$ is $N-2$, and therefore only in $N=4$ is a bivector.
In this case, the dual to the field strength assumes the form of an
inequivalent field strength, 
\begin{eqnarray}
\tilde{F} &=&{\mathbf{i}}F=\dfrac{1}{4}\epsilon ^{\alpha _{1}\alpha
_{2}\beta \gamma }F_{\beta \gamma }\mathbf{e}_{\alpha _{1}}\wedge \mathbf{e}%
_{\alpha _{2}}  \notag \\
&=&\left[ \epsilon ^{0123}F_{01}\mathbf{e}_{2}\wedge \mathbf{e}_{3}+\epsilon
^{0231}F_{02}\mathbf{e}_{3}\wedge \mathbf{e}_{1}+\epsilon ^{0312}F_{03}%
\mathbf{e}_{1}\wedge \mathbf{e}_{2}\right.  \notag \\
&&\mbox{\qquad\qquad}\left. +\epsilon ^{1230}F_{12}\mathbf{e}_{3}\wedge 
\mathbf{e}_{0}+\epsilon ^{2310}F_{23}\mathbf{e}_{1}\wedge \mathbf{e}%
_{0}+\epsilon ^{3120}F_{31}\mathbf{e}_{2}\wedge \mathbf{e}_{0}\right]  \notag
\\
&=&\left[ E_{1}\mathbf{e}_{2}\wedge \mathbf{e}_{3}-E_{2}\mathbf{e}_{1}\wedge 
\mathbf{e}_{3}+E_{3}\mathbf{e}_{1}\wedge \mathbf{e}_{2}\right. \\
&&\mbox{\qquad\qquad}\left. -H_{3}\mathbf{e}_{0}\wedge \mathbf{e}_{3}-H_{1}%
\mathbf{e}_{0}\wedge \mathbf{e}_{1}-H_{2}\mathbf{e}_{0}\wedge \mathbf{e}_{2}%
\right]  \notag \\
&=&\mathbf{e}_{0}\wedge \left( -\mathbf{H}\right) +{\mathbf{i}}_{\left(
space\right) }\mathbf{E},  \label{eqn:0.560}
\end{eqnarray}%
with electric and magnetic vectors exchanged on a one-to-one basis%
\begin{equation}
{\mathbf{E}}\longrightarrow {\mathbf{H}}\mbox{\qquad}{\mathbf{H}}%
\longrightarrow -{\mathbf{E}}.  \label{eqn:0.565}
\end{equation}%
Although (\ref{eqn:0.540}) shows that the duality operation exchanges the
role of the electric-like vector and magnetic-like bivector in any
dimension, it is only in $N=4$ that the bivector can be associated with a
vector in such a way that the dual system can be identified as a transformed
electromagnetic system of the same rank.

The four dimensional Maxwell equations, in the divergence point of view, can
be made symmetric under the duality operation 
\begin{equation}
F\leftrightarrow \tilde{F}\mbox{\qquad\qquad}J^{(e)}\leftrightarrow J^{(m)}
\label{eqn:0.570}
\end{equation}%
by introducing the Dirac magnetic monopole current vector $J^{(m)}$ as the
source for the tensor $\tilde{F}$ in the field equations%
\begin{eqnarray}
d\cdot F &=&-J^{(e)}  \label{eqn:0.580} \\
d\cdot \tilde{F} &=&-J^{(m)}.  \label{eqn:0.590}
\end{eqnarray}%
Since equation (\ref{eqn:0.580}) leads to (\ref{eqn:0.554}) in the form%
\begin{equation}
\nabla \cdot \mathbf{E}=\rho ^{(e)}\mbox{\qquad\qquad}\nabla \times \mathbf{H%
}-\frac{\partial }{\partial t}\mathbf{E}=\mathbf{J}^{(e)}  \label{eqn:0.594}
\end{equation}%
and since $F\rightarrow \tilde{F}$ is equivalent to the exchange (\ref%
{eqn:0.565}), the second field equation (\ref{eqn:0.590}) generalizes (\ref%
{eqn:0.556}) in the form 
\begin{equation}
\nabla \cdot \mathbf{H}=\rho ^{(m)}\mbox{\qquad\qquad}\nabla \times \mathbf{E%
}+\frac{\partial }{\partial t}\mathbf{H}=\mathbf{J}^{(m)}.  \label{eqn:0.596}
\end{equation}%
Then, in the rest frame of a point magnetic source we find the monopole
solution 
\begin{equation}
\mathbf{H}=-\frac{g\mathbf{x}}{\left( \mathbf{x}^{2}\right) ^{3/2}}=-\nabla
\left( -\frac{g}{\left\vert \mathbf{x}\right\vert }\right) .
\end{equation}%
Using (\ref{eqn:0.400}) and (\ref{eqn:0.380}) we can rewrite (\ref{eqn:0.590}%
) in the Jacobi point of view, using 
\begin{equation}
d\cdot \tilde{F}=d\cdot ({\mathbf{i}}F)=(-1)^{N-1}{\mathbf{i}}d\wedge
F=-J^{(m)}
\end{equation}%
which becomes 
\begin{equation}
d\wedge F=-{\mathbf{i}}J^{(m)}.  \label{eqn:0.600}
\end{equation}%
Combining (\ref{eqn:0.600}) and (\ref{eqn:0.580}) Maxwell's equations are 
\begin{equation}
dF=d\cdot F+d\wedge F=-J^{(e)}-{\mathbf{i}}J^{(m)}  \label{eqn:0.620}
\end{equation}%
which is form invariant under the duality transformation induced by the unit
pseudoscalar ${\mathbf{i}}$. Since ${\mathbf{i}}^{2}=-1$ in $N=4$, we may
construct the continuous duality transformation%
\begin{equation}
U\left( \theta \right) =e^{\theta {\mathbf{i}}}=\cos \theta +{\mathbf{i}}%
\sin \theta  \label{eqn:0.630}
\end{equation}%
which acts as%
\begin{eqnarray}
U\left( \theta \right) dF &=&-U\left( \theta \right) \left[ J^{(e)}+{\mathbf{%
i}}J^{(m)}\right] \\
\left[ d\cdot F+d\wedge F\right] \cos \theta +{\mathbf{i}}\left[ d\cdot
F+d\wedge F\right] \sin \theta &=&-\left[ \cos \theta +{\mathbf{i}}\sin
\theta \right] \left[ J^{(e)}+{\mathbf{i}}J^{(m)}\right]
\end{eqnarray}%
so that using (\ref{eqn:0.410}) and (\ref{eqn:0.400}) and separating terms
of equal rank, we obtain%
\begin{eqnarray}
d\cdot F^{\prime } &=&-J^{(e)\prime } \\
d\wedge F^{\prime } &=&-{\mathbf{i}}J^{(m)\prime }
\end{eqnarray}%
where%
\begin{eqnarray}
F^{\prime } &=&F\cos \theta -{\mathbf{i}}F\sin \theta \\
J^{(e)\prime } &=&J^{(e)}\cos \theta -J^{(m)}\sin \theta \\
J^{(m)\prime } &=&J^{(m)}\cos \theta +J^{(e)}\sin \theta .
\end{eqnarray}%
Since $F$ and ${\mathbf{i}}F$ are both bivectors, related by the exchange of
electric and magnetic fields, the transformed field can be identified as an
electromagnetic field. Dirac argued \cite{dirac} that the absence of the
magnetic monopole can be viewed as a convention, according to which we
choose the continuous duality transformation $U\left( \theta \right) $ with
angle $\theta $ that takes $J^{(m)\prime }\rightarrow 0$.

\subsection{Magnetic source in $N>4$ dimensions}

The Maxwell equations were made duality symmetric in $N=4$ by introducing a
source for the second divergence equation in (\ref{eqn:0.590}). Such a
source may be introduced in any dimension, but for $N>4$ the rank of the
dual $\tilde{F}$ is $N-2>2$, so the field equations retain duality
asymmetry. Labeling the rank of multivectors explicitly, the Maxwell
equations become%
\begin{eqnarray}
d\cdot F_{\left( 2\right) } &=&-J_{\left( 1\right) }^{(e)}  \label{eqn:1.640}
\\
d\cdot \tilde{F}_{\left( N-2\right) } &=&-\tilde{J}_{\left( N-3\right)
}^{(m)}  \label{eqn:1.650}
\end{eqnarray}%
where it is convenient to express the magnetic current as%
\begin{equation}
\tilde{J}_{\left( N-3\right) }^{(m)}=\left( -1\right) ^{N-1}{\mathbf{i}}%
J_{\left( 3\right) }^{(m)}.  \label{eqn:1.660}
\end{equation}%
The current trivector $J_{\left( 3\right) }^{(m)}$ has components%
\begin{equation*}
J_{\left( 3\right) }^{(m)}=\tfrac{1}{3!}J_{\left( 3\right) }^{(m)\alpha
\beta \gamma }\mathbf{e}_{\alpha }\wedge \mathbf{e}_{\beta }\wedge \mathbf{e}%
_{\gamma }=\mathbf{e}_{0}\wedge \mathbf{J}_{\left( 2\right) }^{\left(
m\right) }+\rho _{\left( 3\right) }^{(m)}
\end{equation*}%
where the current bivector $\mathbf{J}_{\left( 2\right) }^{\left( m\right) }$
and trivector $\rho _{\left( 3\right) }^{(m)}$ have only space components 
\begin{equation}
\mathbf{J}_{\left( 2\right) }^{\left( m\right) }=\tfrac{1}{2}J_{\left(
2\right) }^{(m)0ij}\mathbf{e}_{i}\wedge \mathbf{e}_{j}\mbox{\qquad}\rho
_{\left( 3\right) }^{(m)}=\tfrac{1}{3!}J_{\left( 3\right) }^{(m)ijk}\mathbf{e%
}_{i}\wedge \mathbf{e}_{j}\wedge \mathbf{e}_{k}\mbox{\qquad}i,j,k=1,\ldots
,N-1,
\end{equation}%
so the dual is%
\begin{equation*}
{\mathbf{i}}J_{\left( 3\right) }^{(m)}=\mathbf{e}_{0}{\mathbf{i}}_{\left(
space\right) }\left( \mathbf{e}_{0}\wedge \mathbf{J}_{\left( 2\right)
}^{\left( m\right) }+\rho _{\left( 3\right) }^{(m)}\right) =\left( -1\right)
^{N}{\mathbf{i}}_{\left( space\right) }\mathbf{J}_{\left( 2\right) }^{\left(
m\right) }+\mathbf{e}_{0}{\mathbf{i}}_{\left( space\right) }\rho _{\left(
3\right) }^{(m)}.
\end{equation*}%
Combining (\ref{eqn:1.640}), which has the non-manifestly covariant form (%
\ref{eqn:0.546a}) and (\ref{eqn:1.650}), the Maxwell equations with electric
and magnetic sources assume the form%
\begin{equation}
\begin{array}{lrr}
\nabla \cdot \mathbf{E}_{\left( 1\right) }=\rho _{\left( 0\right) }^{(e)} & %
\mbox{\hspace{24 pt}} & \left[ \left( -1\right) ^{\frac{\left( N-1\right)
\left( N-2\right) }{2}+N}\right] {\mathbf{i}}_{\left( space\right) }\nabla
\wedge \mathbf{\tilde{F}}_{\left( N-3\right) }-\partial _{0}\mathbf{E}%
_{\left( 1\right) }=\mathbf{J}_{\left( 1\right) }^{(e)}~~%
\end{array}
\label{eqn:1.670}
\end{equation}%
\begin{equation}
\begin{array}{lrr}
\nabla \cdot \mathbf{\tilde{F}}_{\left( N-3\right) }=\left( -1\right) ^{N}{%
\mathbf{i}}_{\left( space\right) }\rho _{\left( 3\right) }^{(m)} & %
\mbox{\hspace{24 pt}} & -{\mathbf{i}}_{\left( space\right) }\nabla \wedge 
\mathbf{E}_{\left( 1\right) }-\partial _{0}\mathbf{\tilde{F}}_{\left(
N-3\right) }={\mathbf{i}}_{\left( space\right) }\mathbf{J}_{\left( 2\right)
}^{\left( m\right) }.%
\end{array}
\label{eqn:1.680}
\end{equation}%
Since 
\begin{equation}
\mathrm{rank}\left[ {\mathbf{i}}_{\left( space\right) }A_{\left( r\right)
}^{\left( space\right) }\right] =\left( N-1\right) -r=3-r  \label{eqn:1.685}
\end{equation}%
in four dimensions, we recover (\ref{eqn:0.594}) and (\ref{eqn:0.596}) by
recalling (\ref{eqn:0.557}) and associating 
\begin{equation}
\mathbf{\tilde{F}}_{\left( 1\right) }=\mathbf{H}\mbox{\qquad\qquad}\rho
_{\left( 0\right) }^{(m)}={\mathbf{i}}_{\left( space\right) }\rho _{\left(
3\right) }^{(m)}\mbox{\qquad\qquad}\mathbf{J}_{\left( 1\right) }^{\left(
m\right) }={\mathbf{i}}_{\left( space\right) }\mathbf{J}_{\left( 2\right)
}^{\left( m\right) }.
\end{equation}%
Since the multivectors in (\ref{eqn:1.640}) and (\ref{eqn:1.650}) are of
different rank when $N\neq 4$, these equations are not duality invariant,
and taking advantage of the Clifford algebraic features of this
representation, it was shown in \cite{duality} that no alternative duality
symmetry exists for this system. It is clear from (\ref{eqn:1.670}) and (\ref%
{eqn:1.680}) that in the general case, exchange of $\mathbf{E}_{\left(
1\right) }$ and $\mathbf{\tilde{F}}_{\left( N-3\right) }$ is not possible
and there is no natural exchange of the electric and magnetic sectors that
would permit consideration of exchange symmetry.

We may put (\ref{eqn:1.640}) and (\ref{eqn:1.650}) into the Jacobi point of
view by applying (\ref{eqn:0.400}) to combine%
\begin{equation}
d\cdot F_{\left( 2\right) }=-J_{\left( 1\right) }^{(e)}\mbox{\qquad\qquad}%
\mathrm{and}\mbox{\qquad\qquad}d\wedge F_{\left( 2\right) }=-J_{\left(
3\right) }^{(m)}
\end{equation}%
as%
\begin{equation}
dF_{\left( 2\right) }=-\left( J_{\left( 1\right) }^{(e)}+J_{\left( 3\right)
}^{(m)}\right) .
\end{equation}%
This system can be made duality symmetric by generalizing the field and
currents to Clifford numbers combining multivectors of appropriate rank, as%
\begin{equation}
F=F_{\left( 2\right) }+G_{\left( N-2\right) }\mbox{\qquad\qquad}J=J_{\left(
1\right) }^{(e)}+J_{\left( 3\right) }^{(m)}+J_{\left( N-3\right)
}^{(e)}+J_{\left( N-1\right) }^{(m)}.  \label{eqn:1.690}
\end{equation}%
The generalized Maxwell equations%
\begin{equation}
dF=-J  \label{eqn:1.695}
\end{equation}%
separate by rank into%
\begin{gather}
d\cdot F_{\left( 2\right) }=-J_{\left( 1\right) }^{(e)}\mbox{\qquad\qquad}%
d\wedge F_{\left( 2\right) }=-J_{\left( 3\right) }^{(m)}  \label{eqn:1.700}
\\
d\cdot G_{\left( N-2\right) }=-J_{\left( N-3\right) }^{(e)}%
\mbox{\qquad\qquad}d\wedge G_{\left( N-2\right) }=-J_{\left( N-1\right)
}^{(m)}  \label{eqn:1.710}
\end{gather}%
and these terms transform under duality according to%
\begin{eqnarray}
{\mathbf{i}}d\cdot F_{\left( 2\right) } &=&\left( -1\right) ^{N-1}d\wedge {%
\mathbf{i}}F_{\left( 2\right) }=-{\mathbf{i}}J_{\left( 1\right) }^{(e)} \\
{\mathbf{i}}d\wedge F_{\left( 2\right) } &=&\left( -1\right) ^{N-1}d\cdot {%
\mathbf{i}}F_{\left( 2\right) }=-{\mathbf{i}}J_{\left( 3\right) }^{(m)} \\
{\mathbf{i}}d\cdot G_{\left( N-2\right) } &=&\left( -1\right) ^{N-1}d\wedge {%
\mathbf{i}}G_{\left( N-2\right) }=-{\mathbf{i}}J_{\left( N-3\right) }^{(e)}
\\
{\mathbf{i}}d\wedge G_{\left( N-2\right) } &=&\left( -1\right) ^{N-1}d\cdot {%
\mathbf{i}}G_{\left( N-2\right) }=-{\mathbf{i}}J_{\left( N-1\right) }^{(m)}.
\end{eqnarray}%
The transformed expressions may be written%
\begin{gather}
d\cdot F_{\left( 2\right) }^{\prime }=-J_{\left( 1\right) }^{(e)\prime }%
\mbox{\qquad}\mbox{\qquad}d\wedge F_{\left( 2\right) }^{\prime }=-J_{\left(
3\right) }^{(m)\prime }  \label{eqn:1.720} \\
d\cdot G_{\left( N-2\right) }^{\prime }=-J_{\left( N-3\right) }^{(e)\prime }%
\mbox{\qquad}\mbox{\qquad}d\wedge G_{\left( N-2\right) }^{\prime
}=-J_{\left( N-1\right) }^{(m)\prime },  \label{eqn:1.730}
\end{gather}%
where%
\begin{gather}
F_{\left( 2\right) }^{\prime }=\left( -1\right) ^{N-1}{\mathbf{i}}G_{\left(
N-2\right) }\mbox{\qquad}\mbox{\qquad}G_{\left( N-2\right) }^{\prime
}=\left( -1\right) ^{N-1}{\mathbf{i}}F_{\left( 2\right) }  \label{eqn:1.740}
\\
J_{\left( 1\right) }^{(e)\prime }={\mathbf{i}}J_{\left( N-1\right) }^{(m)}%
\mbox{\qquad}\mbox{\qquad}J_{\left( N-1\right) }^{(m)\prime }={\mathbf{i}}%
J_{\left( 1\right) }^{(e)}  \label{eqn:1.750} \\
J_{\left( 3\right) }^{(m)\prime }={\mathbf{i}}J_{\left( N-3\right) }^{(e)}%
\mbox{\qquad}\mbox{\qquad}J_{\left( N-3\right) }^{(e)\prime }={\mathbf{i}}%
J_{\left( 3\right) }^{(m)},  \label{eqn:1.760}
\end{gather}%
and duality symmetry is summarized as%
\begin{equation}
dF^{\prime }=-J^{\prime }\mbox{\qquad}F^{\prime }=F_{\left( 2\right)
}^{\prime }+G_{\left( N-2\right) }^{\prime }\mbox{\qquad}J^{\prime
}=J_{\left( 1\right) }^{(e)\prime }+J_{\left( 3\right) }^{(m)\prime
}+J_{\left( N-3\right) }^{(e)\prime }+J_{\left( N-1\right) }^{(m)\prime }.
\end{equation}%
Generalizations of this model were discussed in \cite{duality}.

Writing $G_{\left( N-2\right) }$ and its dual in the non-manifestly
covariant time + space reference frame%
\begin{eqnarray}
G_{\left( N-2\right) } &=&\mathbf{e}_{0}\mathbf{\varepsilon }_{\left(
N-3\right) }+\mathbf{G}_{\left( N-2\right) } \\
{\mathbf{i}}G_{\left( N-2\right) } &=&\mathbf{e}_{0}\wedge \left[ {\mathbf{i}%
}_{\left( space\right) }\mathbf{G}_{\left( N-2\right) }\right] +\left(
-1\right) ^{N}\left[ {\mathbf{i}}_{\left( space\right) }\mathbf{\varepsilon }%
_{\left( N-3\right) }\right] 
\end{eqnarray}%
the duality transformation (\ref{eqn:1.740}) is realized in the exchange of
the noncovariant field sectors%
\begin{eqnarray}
\mathbf{E}_{\left( 1\right) }\leftarrow \mathbf{E}_{\left( 1\right)
}^{\prime }=\left( -1\right) ^{N}\mathbf{\tilde{G}}_{\left( 1\right) } &%
\mbox{\qquad}&\mathbf{\tilde{F}}_{\left( N-3\right) }\leftarrow \mathbf{%
\tilde{F}}_{\left( N-3\right) }^{\prime }=\mathbf{\varepsilon }_{\left(
N-3\right) }  \label{eqn:1.770} \\
\mathbf{\varepsilon }_{\left( N-3\right) }\leftarrow \mathbf{\varepsilon }%
_{\left( N-3\right) }^{\prime }=\mathbf{\tilde{F}}_{\left( N-3\right) } &%
\mbox{\qquad}&\mathbf{\tilde{G}}_{\left( 1\right) }\leftarrow \mathbf{\tilde{%
G}}_{\left( 1\right) }^{\prime }=\left( -1\right) ^{N}\mathbf{E}_{\left(
1\right) }  \label{eqn:1.780}
\end{eqnarray}%
where%
\begin{equation}
\mathbf{\tilde{G}}_{\left( 1\right) }=-{\mathbf{i}}_{\left( space\right) }%
\mathbf{G}_{\left( N-2\right) }\mbox{\qquad}\mathbf{\tilde{F}}_{\left(
N-3\right) }=-{\mathbf{i}}_{\left( space\right) }\mathbf{F}_{\left( 2\right)
}.
\end{equation}%
Equations (\ref{eqn:1.770}) and (\ref{eqn:1.780}) generalize the four
dimensional exchange of the electric and magnetic sectors, but demonstrate
explicitly that the duality transformation does not mix the $\mathbf{E}%
_{\left( 1\right) }$ and $\mathbf{F}_{\left( 2\right) }$ sectors of $%
F_{\left( 2\right) }$ or the $\mathbf{\varepsilon }_{\left( N-3\right) }$
and $\mathbf{G}_{\left( N-2\right) }$ sectors of $G_{\left( N-2\right) }$.
Writing the current $J_{\left( N-1\right) }^{(m)}$as%
\begin{equation}
J_{\left( N-1\right) }^{(m)}=\mathbf{e}_{0}\wedge \mathbf{J}_{\left(
N-2\right) }^{\left( m\right) }+\rho _{\left( N-1\right) }^{(m)}
\end{equation}%
Along with the noncovariant field equations (\ref{eqn:1.670}) and (\ref%
{eqn:1.680}), the covariant expression (\ref{eqn:1.710}) can be written%
\begin{gather}
\nabla \cdot \mathbf{\varepsilon }_{\left( N-3\right) }=\rho _{\left(
N-4\right) }^{(e)}\mbox{\qquad}\left[ \left( -1\right) ^{\frac{\left(
N-1\right) \left( N-2\right) }{2}+N}\right] {\mathbf{i}}_{\left(
space\right) }\nabla \wedge \mathbf{\tilde{G}}_{\left( 1\right) }-\partial
_{0}\mathbf{\varepsilon }_{\left( N-3\right) }=\mathbf{J}_{\left( N-3\right)
}^{\left( e\right) } \\
\nabla \cdot \mathbf{\tilde{G}}_{\left( 1\right) }=\left( -1\right) ^{N}{%
\mathbf{i}}_{\left( space\right) }\rho _{\left( N-1\right) }^{(m)}%
\mbox{\qquad}-{\mathbf{i}}_{\left( space\right) }\nabla \wedge \mathbf{%
\varepsilon }_{\left( N-3\right) }-\partial _{0}\mathbf{\tilde{G}}_{\left(
1\right) }={\mathbf{i}}_{\left( space\right) }\mathbf{J}_{\left( N-2\right)
}^{\left( m\right) }.
\end{gather}%
We observe from equations (\ref{eqn:1.700}) and (\ref{eqn:1.710}) that the
system remains duality symmetric when%
\begin{equation}
J_{\left( 3\right) }^{(m)}=J_{\left( N-3\right) }^{(e)}=0
\end{equation}%
in which case the noncovariant fields $\mathbf{E}_{\left( 1\right) }$
derives from the electric source $J_{\left( 1\right) }^{(e)}$ and $\mathbf{%
\varepsilon }_{\left( N-3\right) }$ derives from the magnetic source $%
J_{\left( N-1\right) }^{(m)}$, while the noncovariant fields $\mathbf{\tilde{%
F}}_{\left( N-3\right) }$ and $\mathbf{\tilde{G}}_{\left( 1\right) }$ remain
sourceless. Thus, in $N>4$ duality symmetry does not guarantee that each
sector of the generalized field equations derives from a source.
Nevertheless, the duality transformation exchanges, according to (\ref%
{eqn:1.770}) and (\ref{eqn:1.780}), an electric solution induced by a
non-zero $J_{\left( 1\right) }^{(e)}$ with a magnetic solution induced by a
non-zero $J_{\left( N-1\right) }^{(m)}$, and may be regarded as a duality
symmetric theory describing both electric and magnetic point sources.
Alternatively, taking%
\begin{equation}
J_{\left( 1\right) }^{(e)}=J_{\left( N-1\right) }^{(m)}=0
\end{equation}%
we find a duality symmetric system in which case the noncovariant field $%
\mathbf{\tilde{F}}_{\left( N-3\right) }$ derives from the magnetic source $%
J_{\left( 3\right) }^{(m)}$ and $\mathbf{\tilde{G}}_{\left( 1\right) }$
derives from the electric source $J_{\left( N-3\right) }^{(e)}$, while the
noncovariant fields $\mathbf{E}_{\left( 1\right) }$ and $\mathbf{\varepsilon 
}_{\left( N-3\right) }$ remain sourceless. In this case, the duality
transformation exchanges an electric solution induced by a non-zero $%
J_{\left( 1\right) }^{(e)}$ with a magnetic solution induced by a non-zero $%
J_{\left( N-1\right) }^{(m)}$, and may also be regarded as a duality
symmetric theory describing both electric and magnetic point sources.

\section{Monopole solution from Lorentz invariance}

B\'{e}rard, Grandati, Lages and Mohrbach \cite{BGLM} studied the Lie algebra
associated with the O(3) invariance of the nonrelativistic system described
in (\ref{eqn:0.170}) to (\ref{eqn:0.210}). Calculating commutation relations
with the angular momentum 
\begin{equation}
L_{i}=m\epsilon _{ijk}x^{i}\dot{x}^{j}  \label{BGLM1}
\end{equation}%
the noncommutivity of the velocities leads to field-dependent terms,%
\begin{eqnarray}
\left[ x_{i},L_{j}\right] &=&-i\hbar \epsilon _{ijk}x_{k}  \label{BGLM2} \\
\left[ \dot{x}_{i},L_{j}\right] &=&-i\hbar \epsilon _{ijk}\dot{x}^{k}+\dfrac{%
i\hbar }{m}\delta _{ij}\left( \mathbf{x}\cdot \mathbf{B}\right) -\dfrac{%
i\hbar }{m}x_{i}B_{j}  \label{BGLM3} \\
\left[ L_{i},L_{j}\right] &=&-i\hbar \epsilon _{ijk}L^{k}-i\hbar \epsilon
_{ijk}x^{k}\left( \mathbf{x}\cdot \mathbf{B}\right) .  \label{BGLM4}
\end{eqnarray}%
Generalizing the operator $\tilde{L}_{i}$ to 
\begin{equation}
\tilde{L}_{i}=L_{i}+Q_{i}  \label{BGLM5}
\end{equation}%
it was shown that the standard angular momentum algebra 
\begin{eqnarray}
\left[ x_{i},\tilde{L}_{j}\right] &=&-i\hbar \epsilon _{ijk}x_{k}
\label{BGLM6} \\
\left[ \dot{x}_{i},\tilde{L}_{j}\right] &=&-i\hbar \epsilon _{ijk}\dot{x}^{k}
\label{BGLM7} \\
\left[ \tilde{L}_{i},\tilde{L}_{j}\right] &=&-i\hbar \epsilon _{ijk}\tilde{L}%
^{k}.  \label{BGLM8}
\end{eqnarray}%
is restored by the choice%
\begin{equation}
Q_{i}=-x_{i}\left( \mathbf{x}\cdot \mathbf{B}\right) ,  \label{BGLM9}
\end{equation}%
when the field $B$ satisfies the structural conditions%
\begin{equation}
x_{j}B_{i}+x_{i}B_{j}+x_{j}x_{k}\partial _{i}B^{k}=0.  \label{BGLM10}
\end{equation}%
Since (\ref{BGLM10}) admits a solution of the form 
\begin{equation}
B_{i}=-\dfrac{gx_{i}}{x^{3}}  \label{BGLM11}
\end{equation}%
the authors argue that the method has led to a magnetic monopole. Moreover,
the total angular momentum $\tilde{L}_{i}$ is conserved under the classical
motion.

In a previous paper \cite{part-I} the B\'{e}rard, Grandati, Lages and
Mohrbach construction was generalized to the full Lorentz group in $N$%
-dimensions. In the notation of the spacetime algebra formalism, the
extended generators%
\begin{equation}
\tilde{M}=M+Q=m\left( x\wedge \dot{x}\right) +x\wedge \left( x\cdot W\right)
-x^{2}W  \label{eqn:2.000}
\end{equation}%
satisfy the canonical Lie algebra for the Lorentz group%
\begin{equation}
\left[ D\cdot x,\tilde{M}\right] =-i\hbar x\wedge D\mbox{\qquad}\left[
D\cdot \dot{x},\tilde{M}\right] =-i\hbar \dot{x}\wedge D
\end{equation}%
when the field is given by 
\begin{equation}
W\left( x\right) =\mathbf{i}\frac{x\wedge U}{\left( x^{2}\right) ^{3/2}}=%
\mathbf{i}G\left( x\right)   \label{eqn:2.100}
\end{equation}%
where $D$ is an arbitrary constant vector, $U$ is a fixed multivector of
rank $N-3$, and the dynamical evolution is restricted to the subspace 
\begin{equation}
x\left( \tau \right) \in x^{U}=\left\{ x~|~x\cdot U=0\right\} .
\label{eqn:2.105}
\end{equation}%
Similarly, the generators satisfy the nearly canonical relations 
\begin{equation}
\left[ \left( D^{\left( 2\right) }\wedge D^{\left( 1\right) }\right) \cdot 
\tilde{M},\tilde{M}\right] =i\hbar \left[ D^{\left( 2\right) }\wedge \left(
D^{\left( 1\right) }\cdot \tilde{M}\right) -D^{\left( 1\right) }\wedge
\left( D^{\left( 2\right) }\cdot \tilde{M}\right) \right] +\Delta _{2}
\end{equation}%
where 
\begin{equation}
\Delta _{2}=i\hbar \frac{\mathbf{i}}{\left( x^{2}\right) ^{1/2}}\left[
\left( D^{\left( 1\right) }\wedge x\right) \wedge \left( D^{\left( 2\right)
}\cdot U\right) -\left( D^{\left( 2\right) }\wedge x\right) \wedge \left(
D^{\left( 1\right) }\cdot U\right) \right] 
\end{equation}%
with arbitrary constant vectors $D^{\left( 1\right) }$, $D^{\left( 2\right) }
$. It was shown that the symmetry-breaking term $\Delta _{2}$ vanishes for
the three generators of O(3) or O(2,1) that leave the subspace $x^{U}$
invariant. Thus, the construction can be extended to any number of
dimensions, but the canonical relations only obtain in a three-dimensional
subspace of the full $N$-dimensional spacetime.

The solution (\ref{eqn:2.100}) may be understood in the following way. In $%
N=4$, a particle moving with uniform timelike four-velocity $U$ produces a
current%
\begin{equation}
J\left( x\right) =\int d\tau \;U~\delta ^{4}\left( x-U\tau \right) 
\end{equation}%
inducing a field that may be found by solving the standard Maxwell equations%
\begin{equation*}
d\cdot F=-J\left( x\right) \mbox{\qquad}d\wedge F=0
\end{equation*}%
via the wave equation%
\begin{equation*}
d^{2}F=-d\wedge J\left( x\right) .
\end{equation*}%
Using the standard Green's function technique, the field is found to be the
Coulomb-like Li\'{e}nard-Wiechert field 
\begin{equation}
F\left( x\right) =-d\wedge \left\{ \frac{1}{2\pi }\int d^{4}x^{\prime
}~J\left( x^{\prime }\right) ~\delta \left[ \left( x-x^{\prime }\right) ^{2}%
\right] \right\} =-d\wedge \frac{U}{\left( x_{\perp }^{2}\right) ^{1/2}}=%
\frac{x\wedge U}{\left( x_{\perp }^{2}\right) ^{3/2}}  \label{eqn:2.110}
\end{equation}%
where the four-velocity in $N=4$ dimensions can be identified with the rank $%
N-3$ multivector $U$. The field $W\left( x\right) $ that satisfies the
requirements for the generator in (\ref{eqn:2.000}) is then the dual to (\ref%
{eqn:2.110}). As shown in section 3.3, the dual $W\left( x\right) =\mathbf{i}%
G\left( x\right) $ in this case is a field tensor with the electric and
magnetic sectors exchanged. The solution $W\left( x\right) $ is seen to
satisfy%
\begin{equation}
d\cdot W=0\mbox{\qquad}d\wedge W=-J\left( x\right)   \label{4-2}
\end{equation}%
which we interpret to describe a magnetic field solution for a magnetic
current, symmetric to the electric Coulomb solution for an electric current.
Thus, in $N=4$, $W\left( x\right) $ corresponds to the standard description
of a magnetic monopole. In particular, by choosing $U=\mathbf{e}_{0}$ along
the time axis, so that the restriction (\ref{eqn:2.105}) becomes%
\begin{equation}
x\left( \tau \right) =\left( 0,\mathbf{x}\right)   \label{4-1}
\end{equation}%
the field $W\left( x\right) $ is 
\begin{equation}
W\left( x\right) =E^{i}\mathbf{e}_{0}\wedge \mathbf{e}_{i}+\tfrac{1}{2}%
\epsilon ^{ijk}B_{i}\mathbf{e}_{j}\wedge \mathbf{e}_{k}=\mathbf{i}\frac{%
\mathbf{e}_{0}\wedge \mathbf{x}}{r^{3}}=-\frac{x^{1}}{r^{3}}\mathbf{e}%
_{2}\wedge \mathbf{e}_{3}+\frac{x^{2}}{r^{3}}\mathbf{e}_{1}\wedge \mathbf{e}%
_{3}-\frac{x^{3}}{r^{3}}\mathbf{e}_{1}\wedge \mathbf{e}_{2}  \label{F3}
\end{equation}%
where%
\begin{equation}
r=\left[ \left( x^{1}\right) ^{2}+\left( x^{2}\right) ^{2}+\left(
x^{3}\right) ^{2}\right] ^{1/2}
\end{equation}%
providing the magnetic monopole solution%
\begin{equation}
\mathbf{E}=0\mbox{\qquad}\mbox{\qquad}\mathbf{B}=-\frac{1}{r^{3}}\left(
x^{1},x^{2},x^{3}\right)   \label{F7}
\end{equation}%
as in (\ref{BGLM11}). In higher dimensions, the multivector $U$ is of rank $%
N-3>1$ and although the field $G\left( x\right) $ retains the Li\'{e}%
nard-Wiechert form, it has rank $N-2>2$. Thus, the fields $G\left( x\right) $
and $W\left( x\right) $ can be identified with the fields $G_{\left(
N-2\right) }$ and $F_{\left( 2\right) }$ given in (\ref{eqn:1.690}). In this
sense, $G_{\left( N-2\right) }$ is an electric-type field satisfying the
equations%
\begin{equation}
d\cdot G_{\left( N-2\right) }=-J_{\left( N-3\right) }\left( x\right) %
\mbox{\qquad}d\wedge G_{\left( N-2\right) }=0
\end{equation}%
and $W\left( x\right) $ is the generalized monopole obtained through the
duality transformation in $N$ dimensions, satisfying%
\begin{equation}
d\cdot W_{\left( 2\right) }=0\mbox{\qquad}d\wedge W_{\left( 2\right)
}=-J_{\left( 3\right) }\left( x\right) .
\end{equation}

We may construct a different kind of solution by replacing the timelike
velocity $U=\mathbf{e}_{0}$ with the spacelike vector $U=\mathbf{e}_{3}$
along the z-axis, characteristic of the relative velocity of an interacting
particle pair. Then, from%
\begin{equation}
F\left( x\right) =\frac{x\wedge \mathbf{e}_{3}}{\rho ^{3}}=\frac{x^{0}}{\rho
^{3}}\mathbf{e}_{0}\wedge \mathbf{e}_{3}+\frac{x^{1}}{\rho ^{3}}\mathbf{e}%
_{1}\wedge \mathbf{e}_{3}+\frac{x^{2}}{\rho ^{3}}\mathbf{e}_{2}\wedge 
\mathbf{e}_{3}  \label{F4}
\end{equation}%
we find the field 
\begin{equation}
W\left( x\right) =E^{i}\mathbf{e}_{0}\wedge \mathbf{e}_{i}+\tfrac{1}{2}%
\epsilon ^{ijk}B_{i}\mathbf{e}_{j}\wedge \mathbf{e}_{k}=\mathbf{i}\frac{%
x\wedge \mathbf{e}_{3}}{\rho ^{3}}=\frac{x^{0}}{\rho ^{3}}\mathbf{e}%
_{1}\wedge \mathbf{e}_{2}+\frac{x^{1}}{\rho ^{3}}\mathbf{e}_{0}\wedge 
\mathbf{e}_{2}-\frac{x^{2}}{\rho ^{3}}\mathbf{e}_{0}\wedge \mathbf{e}_{1},
\label{4d}
\end{equation}%
where%
\begin{equation}
\rho =\left[ -\left( x^{0}\right) ^{2}+\left( x^{1}\right) ^{2}+\left(
x^{2}\right) ^{2}\right] ^{1/2}  \label{F5}
\end{equation}%
generalizes the spatial separation in the action-at-a-distance problems of
nonrelativistic mechanics in the subspace%
\begin{equation}
x=\left( x^{0},x^{1},x^{2},0\right) \in x^{\mathbf{e}_{3}}=\left\{
x~|~x\cdot \mathbf{e}_{3}=0\right\}   \label{F8}
\end{equation}%
invariant under the O(2,1) subgroup of the full Lorentz group. The field
strengths are 
\begin{equation}
\mathbf{E}=\frac{1}{\rho ^{3}}\left( -x^{2},x^{1},0\right) \mbox{\qquad}%
\mathbf{B}=\frac{1}{\rho ^{3}}\left( 0,0,x^{0}\right) .  \label{F6}
\end{equation}%
In this case, closed commutation relations hold among the O(2,1) generators $%
\tilde{M}^{01}$, $\tilde{M}^{02}$, and $\tilde{M}^{12}=\tilde{L}_{3}$, while
the algebra of the generators is broken by field dependent terms for the
boost $\tilde{M}^{03}$ and the rotations $\tilde{M}^{31}=\tilde{L}_{2}$ and $%
\tilde{M}^{23}=\tilde{L}_{1}$.  A description of the relativistic bound
state problem for the scalar hydrogen atom was found \cite{bound} in the
context of the Horwitz-Piron \cite{H-P} formalism, using a potential of this
form. 

\section{Conclusion}

It has been shown \cite{H-S} that the assumption of commutation relations (%
\ref{eqn:0.10}) among quantum position and velocity operators and a second
order force equation (\ref{eqn:0.90}) is sufficiently strong to establish a
Lagrangian system of classical electrodynamics equipped with a canonical
momentum related to velocity through minimum coupling to a vector gauge
field $A^{\mu }$. The Lorentz invariance of this system guarantees
conservation of the generator 
\begin{equation*}
L^{\mu \nu }=x^{\mu }p^{\nu }-x^{\nu }p^{\mu }.
\end{equation*}%
Because the velocity operators do not commute in the presence of an
electromagnetic field, commutation relations with the operator%
\begin{equation*}
M^{\mu \nu }=m\left( x^{\mu }\dot{x}^{\nu }-x^{\nu }\dot{x}^{\mu }\right) 
\end{equation*}%
will contain terms depending on the gauge field, but only in combinations
involving the field strength, obtained as the exterior derivative of the
gauge potential. It was shown in a previous paper \cite{part-I} that the
operators $M^{\mu \nu }$ have a field-dependent extension that satisfies the
canonical commutation relations associated with $L^{\mu \nu }$, when the
field $W^{\mu \nu }$ takes a particular form. Despite the connection to an
underlying $U\left( 1\right) $ gauge theory, the field strength $W^{\mu \nu }
$ was shown in three dimensions \cite{BGLM} and four dimensions \cite{part-I}
to represent a magnetic monopole. In this paper, we examined the solution in 
$N>4$ dimensions, and showed that it represents a magnetic monopole in a
generalized Maxwell theory, in which the Clifford-valued electromagnetic
field strength is constructed to preserved duality symmetry in any dimension.
This notion of duality does not exchange the electric and magnetic sectors
of the noncovariant time + space decomposition of the field strength, but
exchanges among the fields of different rank in a covariant manner. Thus,
the magnetic monopole solution is the dual of an electric field solution of
higher rank. The underlying gauge theory associated with this model will be
discussed in a subsequent paper.

\end{document}